\documentclass[conference]{IEEEtran}
\IEEEoverridecommandlockouts
\usepackage{cite}
\usepackage{amsmath,amssymb,amsfonts}
\usepackage{algorithmic}
\usepackage{graphicx}
\usepackage{textcomp}
\usepackage{xcolor}
\usepackage{bm}
\usepackage{stfloats}
\usepackage{geometry}
\usepackage{fancyhdr}
\def\BibTeX{{\rm B\kern-.05em{\sc i\kern-.025em b}\kern-.08em
    T\kern-.1667em\lower.7ex\hbox{E}\kern-.125emX}}
\begin{document}

\title{A Novel Statistical Analysis Method for Radiation Source Classification}

\author{
\IEEEauthorblockN{
Haobo Geng*,
Yaoyao Li,
Weiping Tong,
Youwei Meng,
Houpu Xiao,
Yicong Liu
}
\IEEEauthorblockA{
\textit{School of Electronics and Information Engineering, Beihang University, Beijing, China}\\
haobogeng@buaa.edu.cn}
}

\maketitle
\renewcommand{\headrulewidth}{0mm}

\begin{abstract}
With the rapid advancement of electronic information technology, the number and variety of unknown radiation sources have increased significantly. Some of these sources share common characteristics, which offers the potential to effectively address the challenge of identifying unknown radiation sources. However, research on the classification of radiation sources remains relatively limited. This paper proposes a big data analysis method that combines linear discriminant analysis (LDA) with a rough neighborhood set (NRS) for radiation source classification, and its effectiveness is validated on the RadioML 2018 dataset. The results indicate that, under certain constraints, all modulation types can be categorized into four distinct classes, laying a foundation for further research on cognitive interference signal cancellation.
\end{abstract}


\section{Introduction}
With the rapid development of electronic information technology, the electromagnetic environment (EME) has become increasingly complex. The number and variety of unknown radiation sources continue to grow, posing challenges to traditional signal analysis. However, classifying these signals is essential for applications such as electronic warfare and communication security.

Although many studies focus on identifying the types of radiation sources using methods such as hypothesis testing [1] and feature engineering [2], modulation schemes often exhibit shared characteristics, suggesting the possibility of grouping them.

This paper combines linear discriminant analysis (LDA) with neighborhood rough set theory (NRS) to analyze radiation source data. Experiments on the RadioML2018.01A dataset [3] demonstrate effective clustering and classification of radiation source signals, providing a foundation for future research on radiation source recognition and interference suppression.
\section{Radiation source classification methods}
\subsection{Selected Feature}
In the field of electromagnetic compatibility (EMC), signal characteristics are typically described in the time, spatial, frequency, energy, and code domains. This paper focuses on target signal features, with spatial propagation effects already reflected in signal parameters such as multipath and fading, which are therefore not considered further. To characterize the temporal dynamics, frequency density, energy fluctuations, and modulation complexity of the electromagnetic environment, the following features are selected [4].
\begin{enumerate}
\item Temporal Dynamics

(a) Temporal Kurtosis: Measures peak sharpness in the time domain.

(b) Singular Spectrum Entropy: Reflects time-series complexity.
\item Frequency Density

(a) Bispectral Integration: Captures amplitude and phase in the frequency domain.

(b) Bandwidth Factor: Ratio of effective bandwidth to central frequency.
\item Energy Fluctuation

(a) Energy Concentration: Indicates energy distribution in the frequency domain.

(b) Fluctuation Index: Quantifies the variation in signal strength over time.
\item Modulation Complexity

(a) Fractal Dimension: Describes detail and complexity across scales.

(b) Lempel-Ziv Complexity: Quantifies the structural complexity of the signal sequence.
\end{enumerate}

Clearly, the features exhibit redundancy, and the complexity of the high-dimensional feature space necessitates dimensionality reduction. To retain the physical significance of the features, a quantitative reduction approach is applied. NRS proves to be an effective tool for reducing continuous data features. This paper utilizes eight features from the RadioML 2018 dataset for subsequent analysis.

\subsection{NRS Reduction Based on LDA}
NRS attribute reduction relies on prior classification results for radiation source signals. This paper first uses LDA for initial dimensionality reduction and classification.  

Also known as fisher linear discriminant, LDA is a supervised technique that maximizes between-class variance while minimizing within-class variance to enhance class separability. Unlike unsupervised methods like principal component analysis, LDA utilizes class labels, making the reduced features better suited for classification tasks [5].

The dataset is ${\bm{S}} = \left\{ {({{\bm{x}}_{\bm{1}}},{y_1}),({{\bm{x}}_{\bm{2}}},{y_2}), \cdots ,({{\bm{x}}_{\bm{m}}},{y_m})} \right\}$, where ${{\bm{x}}_{\bm{i}}}\in {{\mathbb{R}}^{n}}$ represents an arbitrary sample as an $n$-dimensional column vector, and ${{y}_{j}}\in \left\{ {{C}_{1}},{{C}_{2}},\cdots ,{{C}_{K}} \right\}$ denotes the class label of the $j$-th sample, with a total of $K$ classes.

Define ${{\bm{\mu }}_{\bm{j}}}$ as the mean vector of the $j$-th class samples.
\begin{equation}
    {{\bm{\mu }}_{\bm{j}}} = \frac{1}{{{N_j}}}\sum\limits_{{{\bm{x}}_{\bm{i}}} \in {{\bm{S}}_{\bm{j}}}} {{{\bm{x}}_{\bm{i}}}} ,{\bm{ }}j = 1,2, \cdots ,K
\end{equation}
where ${{N}_{j}}$ denotes the number of samples in the $j$-th class, and ${{\bm{S}}_{\bm{j}}}$ represents the sample set of the $j$-th class.

The within-class scatter matrix measures the dispersion of samples within the same class, reflecting how samples are distributed around the mean.
\begin{equation}
   {{\bm{S}}_{\bm{W}}} = \sum\limits_{j = 1}^K {\sum\limits_{{{\bm{x}}_{\bm{i}}} \in {{\bm{S}}_{\bm{j}}}} {{{({{\bm{x}}_{\bm{i}}} - {{\bm{\mu }}_{\bm{j}}})}^T}({{\bm{x}}_{\bm{i}}} - {{\bm{\mu }}_{\bm{j}}})}}
\end{equation}

The between-class scatter matrix measures the dispersion between samples of different classes, representing the distances between the centroids of different classes. 
\begin{equation}
  {{\bm{S}}_{\bm{B}}} = \sum\limits_{j = 1}^K {{N_j}{{({{\bm{\mu }}_{\bm{j}}} - {\bm{\mu }})}^T}({{\bm{\mu }}_{\bm{j}}} - {\bm{\mu }})}
\end{equation}
where ${\bm{\mu }}$ is the overall mean vector of all samples.

LDA aims to find the optimal projection direction by maximizing the ratio of the between-class to within-class scatter matrices. The multi-class objective function is
\begin{equation}
\underbrace {\arg {\bm{ }}\max }_{\bm{W}}{\bm{ }}J({\bm{W}}) = \frac{{\mathop \prod \limits_{diag} {\mkern 1mu} {{\bm{W}}^T}{{\bm{S}}_{\bm{B}}}{\bm{W}}}}{{\mathop \prod \limits_{diag} {\mkern 1mu} {{\bm{W}}^T}{{\bm{S}}_{\bm{W}}}{\bm{W}}}}
\end{equation}
where $\underbrace {\arg {\bm{ }}\max }_{\bm{W}}$ represents finding a ${\bm{W}}$ that maximizes $J({\bm{W}})$, $\mathop \prod \limits_{diag} {\bm{A}}$ denotes the product of the diagonal elements of ${\bm{A}}$, and ${\bm{W}}$ is the projection matrix.

The maximum value of (4) is the largest eigenvalue of ${\bm{S}}_{\bm{W}}^{ - 1}{{\bm{S}}_{\bm{B}}}$. The matrix ${\bm{W}}$ consists of the eigenvectors corresponding to the largest $d$ eigenvalues of ${\bm{S}}_{\bm{W}}^{ - 1}{{\bm{S}}_{\bm{B}}}$, given by
\begin{equation}
{\bm{W}} = eig({\bm{S}}_{\bm{W}}^{ - 1}{{\bm{S}}_{\bm{B}}})
\end{equation}
where $eig()$ calculates eigenvalues and eigenvectors. For $d$-dimensional reduction, the top $d$ eigenvectors are selected as the projection directions.

After dimensionality reduction, each sample is transformed as ${{\bm{z}}_{\bm{i}}} = {{\bm{W}}^T}{{\bm{x}}_{\bm{i}}}$. And the output dataset is ${{\bm{S}}'} = \left\{ {({{\bm{z}}_{\bm{1}}},{y_1}),({{\bm{z}}_{\bm{2}}},{y_2}), \cdots ,({{\bm{z}}_{\bm{m}}},{y_m})} \right\}$. The reduced result retains the most relevant information for classification, providing reliable prior information for attribute reduction in neighborhood rough sets.

Next, we introduce NRS, an extension of traditional rough sets. Unlike precise partitioning, NRS uses "neighborhoods" to address data ambiguity, grouping similar samples [6]. The feature vectors and class labels from the previous section correspond to the conditional and decision attributes, respectively.

Let $NDS=(U,R,V,f)$ be an information system, where $U=\{{{z}_{1}},{{z}_{2}},\cdots ,{{z}_{m}}\}$ is the universe of objects, $R=C\bigcup D$ is the attribute set with condition attributes $C$ and decision attributes $D$, and $V=\bigcup {{V}_{a}}$ represents the attribute value set, where ${{V}_{a}}$ is the range of attribute $a$. The information function $f:U\times R\to V$ assigns values such that $\forall a\in R,\text{ }z\in U$, $f(z,a)\in {{V}_{a}}$.

Specifically, for a neighborhood rough set decision system, given ${{z}{i}}\in U$ and a subset of condition attributes $B\subseteq C$, the neighborhood of ${{z}{i}}$ under $B$ is defined as
\begin{equation}
{\delta _B}({z_i}) = \left\{ {{z_j}|{z_j} \in U,{\Delta ^B}({z_i},{z_j}) \le \delta } \right\}
\end{equation}
where $\Delta$ is defined as $\Delta ({{z}_{i}},{{z}_{j}})={{\left[ \sum\limits_{k=1}^{n}{{{\left( {{z}_{ik}}-{{z}_{jk}} \right)}^{2}}} \right]}^{\frac{1}{2}}}$, where ${{z}_{ik}}$ and ${{z}_{jk}}$ represent the $k$-th attribute values of objects ${{z}_{i}}$ and ${{z}_{j}}$, respectively, and $n$ is the total number of condition attributes. $\Delta ({{z}_{i}},{{z}_{j}})$ represents the Euclidean distance between ${{z}_{i}}$ and ${{z}_{j}}$, while $\delta$ denotes the neighborhood radius.

Let $Z$ be a subset of $U$. For $\forall B\subseteq C$, the lower approximation of $Z$ in neighborhood space is defined as
\begin{equation}
\underline {apr_B} (Z) = \left\{ {{z_i}|{\delta _B}({z_i}) \subseteq Z,{z_i} \in U} \right\}
\end{equation}

For the NRS decision system $NDS=(U,R,K)$, if $D$ partitions $U$ into $K$ equivalence classes (${Z}{1},{Z}{2},\cdots ,{Z}_{K}$), the lower approximation of $D$ with respect to $B$ is defined as
\begin{equation}
\underline {ap{r_B}} (D) = \mathop  \cup \limits_{i = 1}^K \underline {ap{r_B}} ({Z_i})
\end{equation}

The dependency degree of $D$ on the conditional attribute subset $B$ is defined as
\begin{equation}
{\gamma _B}(D) = \frac{{\left| {\underline {ap{r_B}} (D)} \right|}}{{\left| U \right|}}
\end{equation}

If $a \in (C - B)$, the significance of $a$ relative to $B$ is
\begin{equation}
SIG(a,B,D) = {\gamma _{B \cup a}}(D) - {\gamma _B}(D)
\end{equation}

This achieves the most critical application of rough set theory: knowledge extraction and attribute reduction, thereby avoiding dimensional explosion. When $B$ satisfies $\forall a\in B,\text{ }{{\gamma }_{B-a}}(D)<{{\gamma }_{B}}(D)$ and ${{\gamma }_{B}}(D)={{\gamma }_{C}}(D)$, $B$ is called the minimal subset of attributes.

\section{Experimental verification}
\subsection{Public Dataset}
This paper uses the RadioML 2018 dataset, a benchmark in radio signal processing, covering 24 modulation types with simulated channel effects. From this dataset, 1,000 samples with varying signal-to-noise ratios were randomly selected for each modulation type, totaling 24,000 samples for analysis.
\subsection{Validation Process}
To reduce external interference and improve data reliability, power normalization is applied to the complex signal $z(n)$ with $N$ sampling points, minimizing the effect of signal strength on classification accuracy.
\begin{equation}
\overline z \left( n \right) = \frac{{z\left( n \right)}}{{\sqrt {\frac{1}{N}\sum\limits_{n = 1}^N {{{\left| {z\left( n \right)} \right|}^2}} } }}
\end{equation}

Using Section II.A’s definitions, eight features were computed to form the dataset ${\bm{S}}$. Modulation schemes of the same type but different orders share similar principles, leading to common post-processing methods across various domains. Thus, in LDA, ${\bm{S}}$ groups modulation schemes of the same type but different orders into a single initial category, as shown in Table I.
\begin{figure}[htpb]
      \centering
      \includegraphics[scale=1]{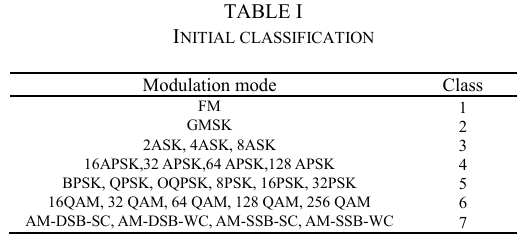}
\end{figure}

The ${\bm{W}}$ was computed using (5). To evaluate the impact of dimensionality $d$ on accuracy, ten-fold cross-validation was performed, and the results are shown in Fig. 1.
\begin{figure}[htpb]
      \centering
      \includegraphics[scale=0.45]{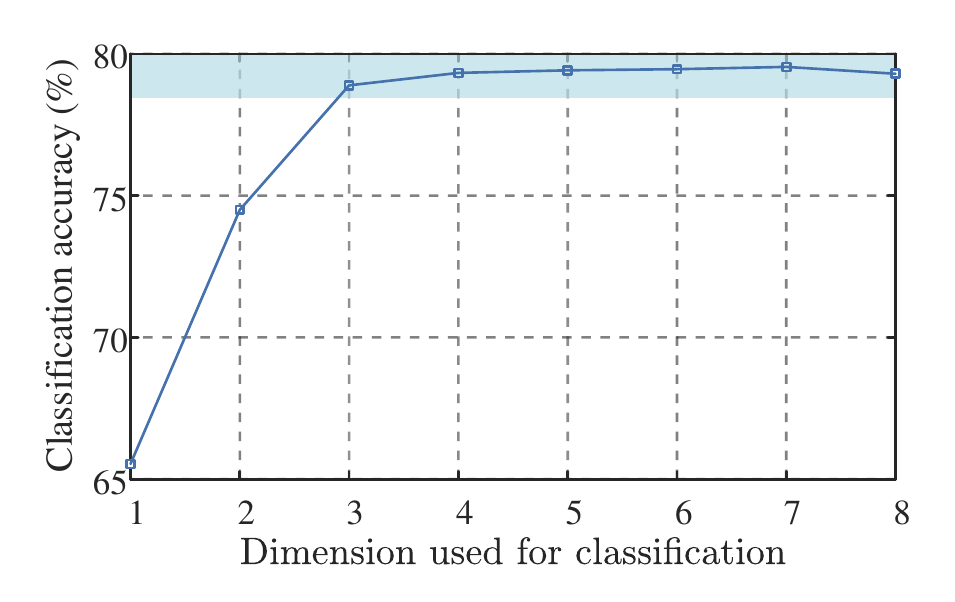}
      \caption{Classification accuracy varies with the dimensionality.}
      \label{fig1}
\end{figure}

From Fig. 1, it can be seen that the classification accuracy based on the original dataset is 79.3\%. When $d=3$, the accuracy approaches the ideal level. Therefore, we choose the 3-dimensional projection matrix, and the dimensionality-reduced results are visualized in Fig. 2. Based on this clustering result, the class attributes are updated accordingly.
\begin{figure}[htpb]
      \centering
      \includegraphics[scale=0.6]{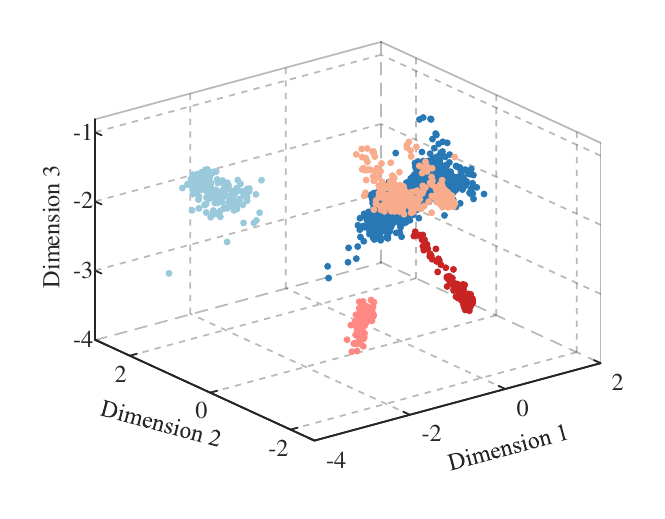}
      \caption{Visualization of samples reduced to three dimensions using LDA.}
      \label{fig2}
\end{figure}

Since LDA relies on 8 features, which is computationally intensive and redundant, we reduce the features using NRS based on LDA's classification results. The reduction depends on the neighborhood radius $\delta$, so we conduct experiments under different $\delta$ values. The importance distribution of the 8 features is shown in Fig. 3.
\begin{figure}[htpb]
      \centering
      \includegraphics[scale=0.5]{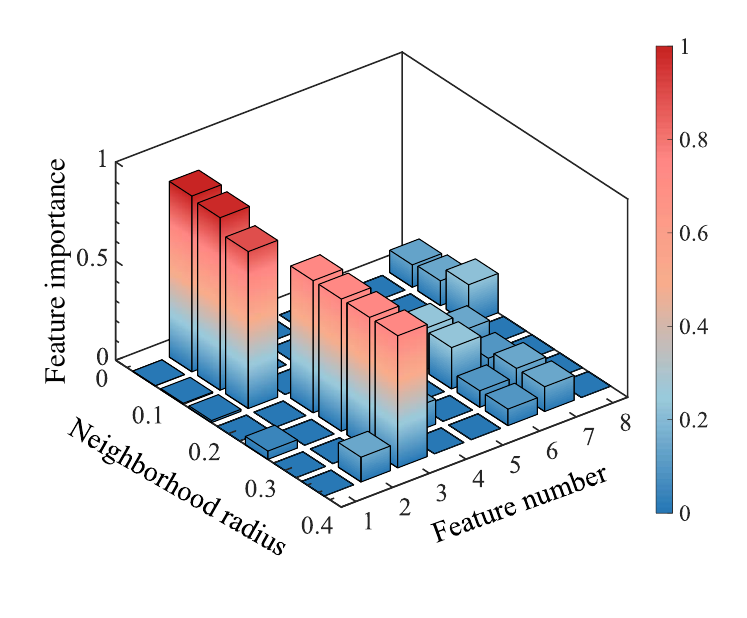}
      \caption{The importance calculation results of the 8 features under various neighborhood radius conditions.}
      \label{fig3}
\end{figure}

Based on the reduction results under various $\delta$ conditions, the second and third attributes, namely singular spectral entropy and bispectral integral, are selected as the final classification features. The updated classification results are shown in Fig. 4, with detailed classifications in Table II.
\begin{figure}[htpb]
      \centering
      \includegraphics[scale=0.45]{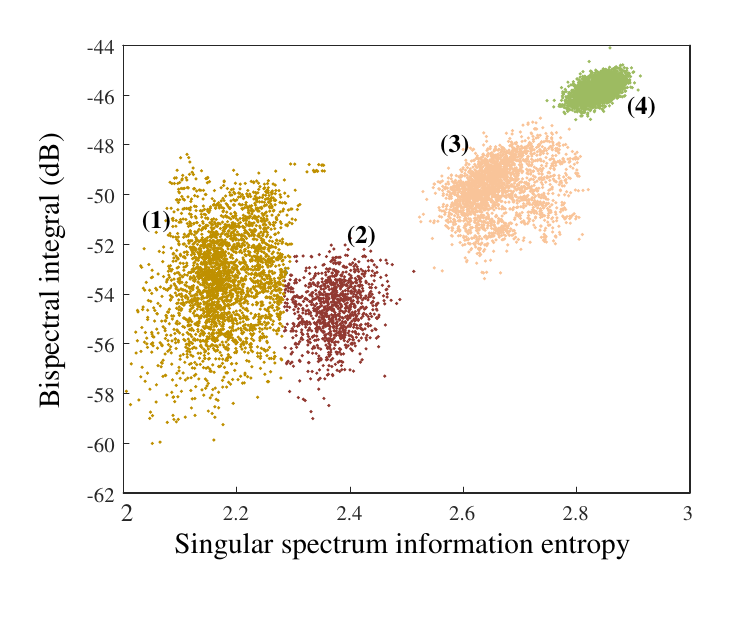}
      \caption{The classification results of the samples based on the reduced features.}
      \label{fig4}
\end{figure}

\begin{figure}[htpb]
      \centering
      \includegraphics[scale=1]{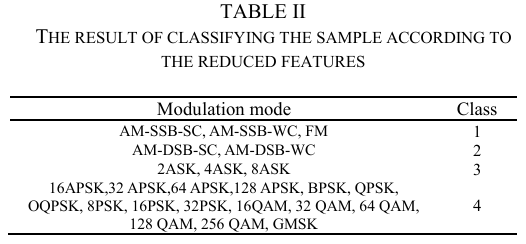}
\end{figure}

Thus, we have developed an understanding of signal types based on feature engineering and statistical analysis. Future research will build upon this work to further explore cognitive interference signal cancellation.

\section{Conclusion}
This paper conducts an in-depth analysis of signal features using the RadioML 2018 dataset and proposes a simplified radiation source classification method that integrates linear discriminant analysis with neighborhood rough set theory. This study offers novel technical support for interference cancellation in electronic warfare and communication security, demonstrating both substantial theoretical significance and practical application potential.

\section*{Acknowledgment}
This work was supported by the National Key Research and Development Program under Grant 2023YFB3306900.

\end{document}